# A Guide for Material and Design Choices for Electro-Optic Modulators and recent 2D-Material Silicon Modulator Demonstrations

**Rubab Amin[1], Mario Zhizhen[1], Rishi Maiti[1], Mario Miscuglio[1], Hamed Dalir[3], Jacob B. Khurgin[2], Volker J. Sorger[1],***

[1]Department of Electrical and Computer Engineering, George Washington University, Washington, DC 20052, USA
[2]Department of Electrical and Computer Engineering, Johns Hopkins University, Baltimore, Maryland 21218, USA
[3]Omega Optics, Inc. 8500 Shoal Creek Blvd., Bldg. 4, Suite 200, Austin, Texas 78757, USA
*Author email address: sorger@gwu.edu

**Abstract:**
**Electro-optic modulation performs a technological relevant functionality such as for communication, beam steering, or neuromorphic computing through providing the nonlinear activation function of a perceptron. Wile Silicon photonics enabled the integration and hence miniaturization of optoelectronic devices, the weak electro-optic performance of Silicon renders these modulators to be bulky and power-hungry compared to a single switch functionality known from electronics. To gain deeper insights into the physics and operation of modulators hetero-generous integration of emerging electro-optically active materials could enable separating light passive and low-loss light routing from active light manipulation. Here we discuss and review our recent work on a) fundamental performance vectors of electro-optic modulators, and b) showcase recent development of heterogeneous-integrated emerging EO materials into Si-photonics to include an ITO-based MZM, a Graphene hybrid-plasmon and the first TMD-MRR modulator using a microring resonator. Our results indicate a viable path for energy efficient and compact Silicon photonic based modulators.**

## 1. Introduction

Efficient electro-optic modulation and light detection is important in order to reduce the power consumption for data communication interconnects, but also for emerging processors such as neuromorphic photonic computing [1,2]. For instance in the latter, they can perform the nonlinear activation function of an analog photonic perceptron, with the advantage of not limited by an digital processor's nanosecond-slow clock speed, and the physical dimensions of the modulator can be designed to ensure a particular signal-to-noise ratio with the neural network, which is relevant for the neural network's cascadability [2]. Heterogeneous integration of emerging materials such as TMDs allows to take advantage of the low losses of Si/SiN waveguides and cavities, and combines this with efficient electro-optic responses of low-dimensional materials showing stronger responses to electric field stimuli due to the reduced dimension and thus lower coulomb screening than bulk materials [3].

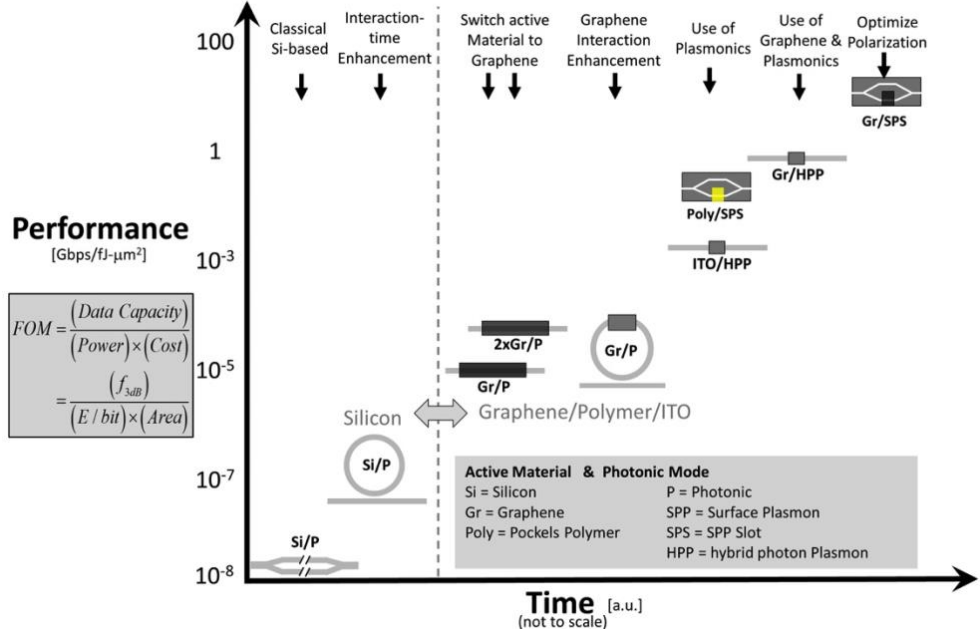

**Figure 1.** The roadmap of EOM development using the pre-defined performance matrix. Graphene showed predominant advantage compared to silicon as the active material selection, while plasmonic mode (HPP/SPP/SPS) provide a much higher light-active material interaction by squeezing the mode into the active material region. Thus the integration of unity index change material and plasmonic mode could reduce the device size and bring the performance to the upper side sweet spot [4].

We start our discussion by summarizing selected studies on EOM development focusing on active material selection and photonic platform design using a unified performance benchmark, where we define the figure-of-merit (FOM) as $FOM = \frac{Data\ capacity}{Power \times Cost} = \frac{f_{3-dB}}{(E/bit) \times Area}$ [4]. For seamless integration with CMOS technology, extensive investigation has been sought to utilize the classic semiconductor, silicon, for electro-optic modulation by using free carrier injection/depletion. However, due to the intrinsically low carrier concentration, and weak free carrier dispersion of silicon, either a long arm Mach-Zehnder interferometer (MZI) or a low loss, high-$Q$ resonator is needed to increase the light-matter interaction (LMI). Since the RC-delay of the device and power consumption both suffer from larger device volume, pioneering work on silicon MZI and micro-ring modulators have a rather low overall FOM, which sets our baseline for the performance matrix (**Fig. 1**) [3]. With the emergence of low dimensional materials, graphene has shown high tunability as an absorptive material. The silicon waveguide used is rather compact (tens of micrometers), benefiting from the drastic change of graphene's absorption compared to silicon. However, although the waveguide dimension was optimized to maximize the mode overlap between the propagating mode and the active material (graphene), at telecom wavelength, the conventional photonic mode still has a certain cutoff physical dimension (220 nm for Si at 1550 nm wavelength), which is usually much larger compare to the 0.34 nm thickness of single atom layer active material graphene. Thus, for either graphene EOM work from 2011 or 2012 in Fig. 1 the device dimension is still bulky due to the low LMI. Nevertheless, in 2015 the Lipson-group demonstrated a graphene modulator with 30 GHz-fast modulation. Yet, this device still suffers from the bulky dimension and reduced thermal stability from the ring resonator. On the other hand, increased optical confinement has been explored such as in plasmonic modulators. Here, the plasmonic modes increases the modal overlap factor of the active material with that of the optical mode, and provides a free metallic contact which has a smaller series resistance in the device, hence reducing the RC delay compared to the photonic platform which usually requires doped semiconductors as the electrical contact. By squeezing the mode into the active material region to increase the LMI and thus the FOM, modulators have been demonstrated by using other active materials such as transparent conductive oxides (e.g. ITO), but also Pockels-effect field-driven (non charge-drive) materials such as polymers. Following this hybrid-integration scheme (e.g. silicon plus novel-material) Leuthold's group has shown that the latter enables a platform for ultra-fast modulation approaching 200 GHz.

## 2. Modulator Material and Mode Considerations

Electro-optic modulation performs the conversion between the electrical and optical domain with applications in data communication for optical interconnects, but also for novel optical computing algorithms such as providing nonlinearity at the output stage of optical perceptrons in neuromorphic analog optical computing. While resembling an optical transistor, the weak light-matter-interaction makes modulators $10_5$ times larger compared to their electronic counterparts. Since the clock frequency for photonics on-chip has a power-overhead sweet-spot around tens of GHz, ultrafast modulation may only be required in long-distance communication, but not for short on-chip links. Hence, the search is open for power-efficient on-chip modulators beyond the solutions offered by integrated photonic-foundries to date. In this context, we will discuss a) fundamental scaling vectors towards attojoule per bit efficient modulators on-chip, along with b) selected recent experimental demonstrations of novel plasmonic modulators with near and sub-fJ/bit efficiencies (**Fig. 2**) [5]. As aforementioned, the important performance matrix for EOMs include modulation speed, power consumption, and (less importantly) the footprint area (possibly 3D volume).

Here we focus on carrier-based electro-absorption models which are implemented via capacitive gating (**Fig. 2a**). The resulting imaginary index change is relatively smaller the spectral detuning form the material 'transition', such as the exciton at 2-dimensional (2D) materials such as the group of transition-metal-dichalcogenides (TMD). A key question is how 'sharp' the modulation transition can be with respect to the gate voltage, where smaller voltage leads to lower energy dissipation and more efficient designs [5]. This transition sharpness for all modulator types depends on the amount of broadening in the material system such as (in)homogeneous effects and naturally temperature smearing ($\sim k_B T$) (**Fig. 2b**). A detailed analysis of the required voltage to evoke a certain amount of optical absorption shows that 2-level carrier-blocking modualtors such as found in quantum dots result in very steep switching devices, whereas the strongly-bound exciton in 2D TMDs makes them stable against modulation (**Fig. 2c**). Graphene and quantum well-based modulators such as those based on the quantum confined Stark effect show a performance between these two extremes. Interestingly, for the aforementioned photonic neural networks the latter exhibit the best platform for nonlinear activation due to their pronounced sigmoidal S-shape behavior [2]. Subsequent performance parameters such as capacitor size, required voltage (here to obtain 10dB signal modulation), and energy-per-bit and 3dB bandwidth (speed), can be derived and are derivatives from the transfer function [4, 5].

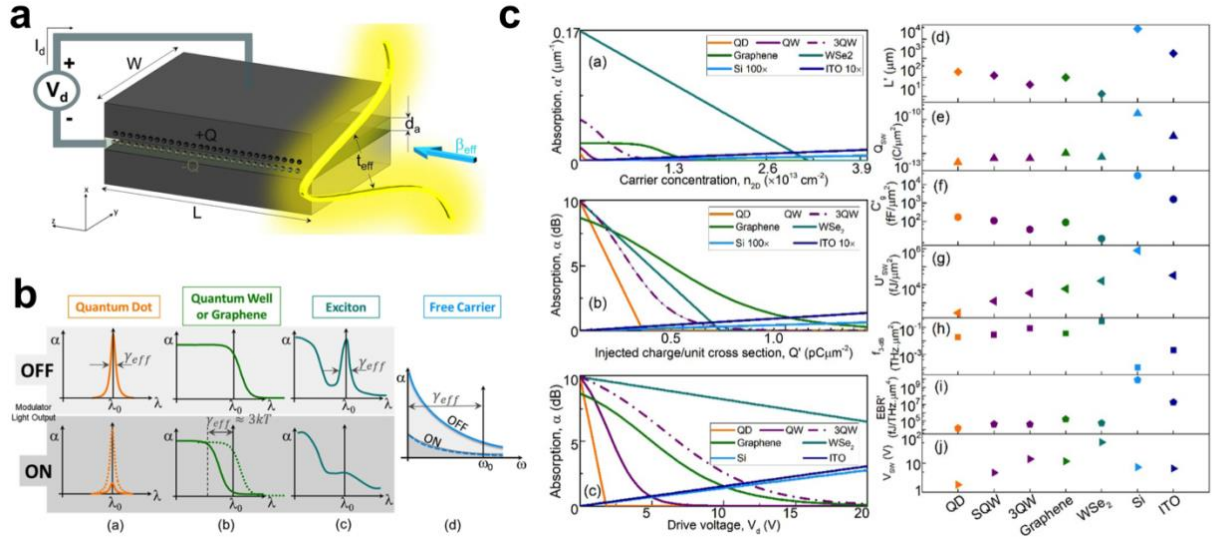

**Figure 2. Comprehensive ab-initio analysis for EAMs. a.** We take a schematic of a waveguide and define the modal overlap with the active material as $t_{eff}$. **b.** Then we take 4 different material modulation mechanisms by focusing on electro-absorption materials (e.g. charge driven). **c.** We then analyze the performance such as absorption vs. drive voltage, speed, energy/bit and the FOM energy-bandwidth-ratio. We find that this fundamental modulator performance is relatively flat for all 2-level blocking mechanisms (state-filling, Pauli-blocking, exciton modulation), but much weaker for free-carrier-based devices (e.g. Si or ITO) [5]. $\lambda$ = 1550 nm, gate oxide = 100 nm.

## 3. Recent Heterogeneous Si-photonic-integrated Modulator Demonstrations

Electro-optic modulators transform electronic signals into the optical domain and are critical components in modern telecommunication networks, RF photonics, and emerging applications in quantum photonics, neuromorphic photonics, and beam steering. All these applications require integrated and voltage-efficient modulator solutions with compact form factors that are seamlessly integrable with Silicon photonics platforms and feature near-CMOS material processing synergies [6]. However, existing integrated modulators are challenged to meet these requirements. Conversely, emerging electro-optic materials heterogeneously and monolithically integrated with Si photonics open up a new avenue for device engineering. Indium tin oxide (ITO) is one such compelling material for heterogeneous integration in Si exhibiting formidable electro-optic effect characterized by unity-order index change at telecommunication frequencies. Here we overcome these limitations and demonstrate a monolithically integrated ITO electro-optic modulator based on a Mach Zehnder interferometer (MZI) featuring a high-performance half-wave voltage and active device length product of $V\pi L = 0.52$ V·mm (**Fig. 3**) [7]. We show, that the unity-strong index change enables a 30 micrometer-short π-phase shifter operating ITO in the index-dominated region away from the epsilon-bear-zero ENZ point for reduced losses. This device experimentally confirms electrical phase shifting in ITO enabling its use in applications such as dense on-chip communication networks, nonlinearity for activation functions in photonic neural networks, and phased array applications for LiDAR

Lastly, we report on a novel electro-optic modulator based on heterogeneous integration of TMDs onto a silicon photonics microring resonator (MRR) (**Fig. 4**). To our knowledge this is the first time that either device is demonstrated in such a platform, however Lipson's Group recently demonstrated strong phase-

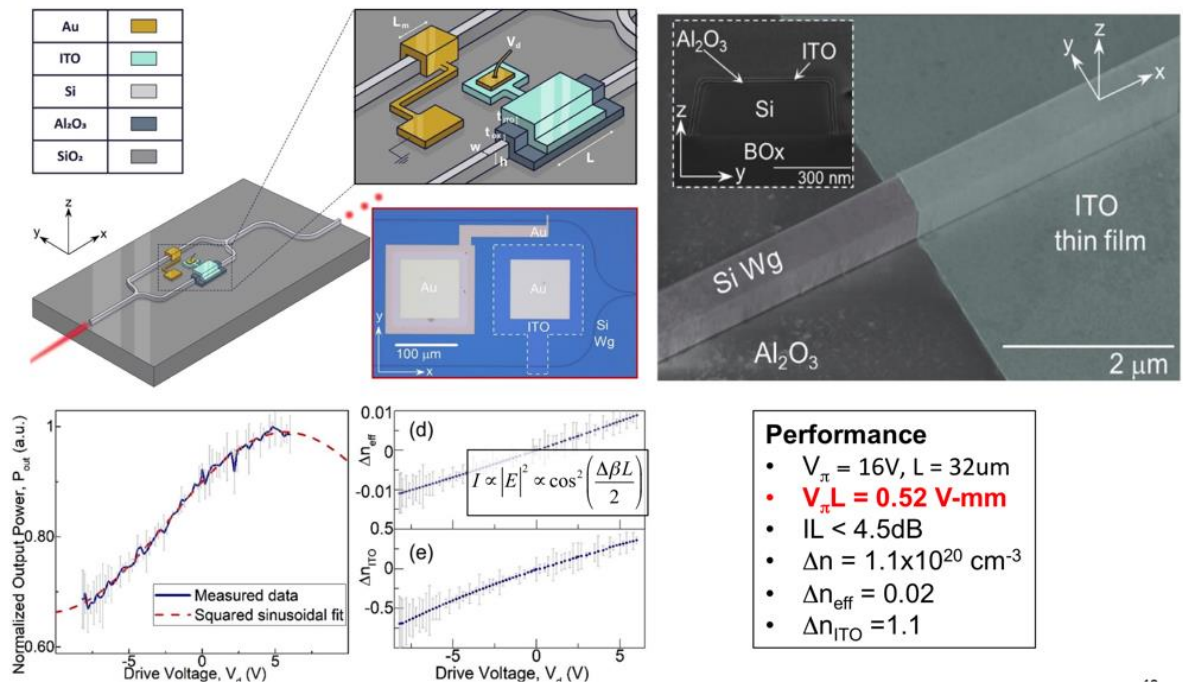

**Figure 3. First demonstration of an ITO-based Silicon MZI modulator.** The device shows a high performance of $V\pi L = 0.52$V-mm. Note, this is not an ENZ device, since ENZ bears too much loss. In a phase shifter we aim to stay away from ENZ to maximize the dn/dκ, as realized here. Fitting the optical power change with voltage shows a modal index change of 0.02 corresponding to active ITO index change of beyond unity (1.1) [7]. λ = 1550 nm.

modulation in an MZI structure [8]. The EOM is realized in a two-terminal in-plane electrode configuration where 2D hBN flakes are used as gate dielectric, and MoS$_2$ as the actively gated material (**Fig. 4a**) [9-11]. Given the wavelength of 1550 nm, we are far away form the exciton resonance of the TMD. Since $dn/d\lambda$ decays slower than $d\kappa/d\lambda$, find a basically unchanged Q-factor with bias. The device shows a ~2dB modulation for 4V of applied bias (**Fig. 4b**). The effect seems rather weak at first, however the gating 'oxide' is the lateral offset of center-ring electrode which is about 1 µm away from the Silicon microring (**inset, Fig. 4b**). Normalizing the electrostatics to an equivalent gate oxide of 10 nm, this would result in 20mV/dB of modulation and would constitute a rather efficient device. To expand upon this, the top electrode is placed inside the MRR in such a way to not overlap with the MRR; thus the optical mode of the device is photonic, and not plasmonic. The electric field lines, are therefore horizontal (parallel) to the substrate surface and thus in-plane with the TMD matching the polarization of the waveguide. This ensures an optimized overlap of the applied gate voltage with the optical mode inside the silicon waveguide of the MRR. The index change of the effective optical mode is proportional to the mode-overlap, and thus ensures

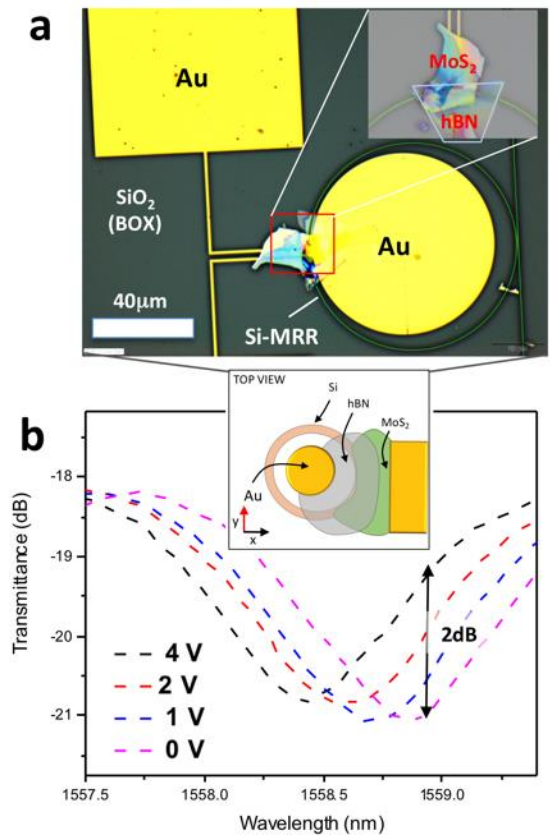

**Figure 4.** Two novel optoelectronic device demonstrations, namely a electro-optic modulator and a photodetector based on heterogeneous integration of TMDs onto a silicon photonics microring resonator (MRR). **a&b,** Optical image and performance of the electro-optic (phase shifting) modulator. TMD = MoS$_2$, Epi SOI silicon = 220nm, BOX = 1µm, $\lambda$ = 1550nm, showing 2dB of modulation for 4V. The device is laterally gated and the optical device-section mode is photonic.

efficient modulation.

**Application Discussion**

Modulators are universal building blocks for i) data links and interconnects including hybrid-technology options [12] including dual-function optoelectronic devices [13], ii) photonic integrated network-on-chip technology [14], iii) photonic digital-to-analog converters [15], and photonic analog processors such as iv) photonic residue arithmetic processors relying on cross-bar routers [16, 17], v) analog partial differential equation solvers [18], vi) photonic integrated neural networks [19-22], vii) photonic reprogrammable circuits such as content-addressable memories [23], or for optical phase-arrays in beam steering [24]. We note, that two dimensional materials, such as graphene can be grown in a foundry-near method such as using chemical vapor deposition (CVD) and subsequent etching. could be deployed for high-speed while preserving high modulation voltage efficiency showing <fJ/bit modulation efficiency [39] and the potential for 100 GHz fast modulation [40]. Scaling laws for opto-electronics and cavity options could be used further to reduce energy consumption of next generation 2D material based modulators [41, 42].

**5. Conclusions**

We discussed electro-absorption based physical effects and interdependencies effective modulator performance. Here we find that the material broadening effects are mainly determining the voltage-gating efficiency, which will thence determine other modulator-relevant performance parameters such as speed and energy-efficiency. Then we discuss two modulator demonstrations utilizing heterogeneous integration for one device on ITO for the other a 2D TMD material onto silicon photonics. For the former we find a performance $V\pi L$ = 0.52 V-mm, and for the latter a modulation of an equivalent of 20 mV/dB, which is near the noise-floor at room temperature. This short review highlights that there is still room for unexplored work in the field of electro-absorption modulators, and suggests, in addition to data communication, photonic neural networks as a possible application for these devices by realizing that the modulator's transfer function maps synergistically onto the nonlinear activation function of a photonic perceptron.

**6. Acknowledgements**

Air Force Office of Scientific Research (AFOSR) (FA9550-17-1-0377, FA9550-17-P-0014); Army Research Office (ARO) (W911NF-16-2-0194).

## 7. References

1. M. Miscuglio, A. Mehrabian, Z. Hu, S. I. Azzam, J. K. George, A. V. Kildishev, M. Pelton, V. J. Sorger, "All-optical Nonlinear Activation Function for Photonic Neural Networks", Optical Material Express 8(12), 3851-3863. (2018)
2. J. K. George, R. Armin, B. Shastri, P. Prucnal, T. El-Ghazawi, V. J. Sorger, "Noise and Nonlinearity of Electro-optic Activation Functions in Neuromorphic Compute Systems" *arXiv* preprint: 1809.03545 (2018)
3. R. Amin, et al. "Active Material, Optical Mode and Cavity Impact on electro-optic Modulation Performance" *Nanophotonics* 7, 455 (2017)
4. R. Amin, Z. Ma, R. Maiti, S. Khan, J. B. Khurgin, et al., "Attojoule-Efficient Graphene Optical Modulators", *Appl. Opt.*, 57, 18, 1-11. (2018)
5. R. Amin, J. B. Khurgin, V. J. Sorger, "Waveguide based Electroabsorption Modulator Performance", *Opt. Exp.*, 26, 12, 15445-15470 (2018).
6. V. J. Sorger, R. Amin, J. B. Khurgin, Z. Ma, S. Khan, "Scaling Vectors for Atto-Joule per Bit Modulators" *J. Optics* 20, 014012 (2018).
7. R. Amin, R. Maiti, C. Carfano, Z. Ma, M. H. Tahersima, Y. Lilach, D. Ratnayake, H. Dalir, V. J. Sorger, "0.52 V-mm ITO-based Mach- Zehnder Modulator in Silicon Photonics", APL Photonics, 3,12. (2018)
8. Datta, S.H. Chae, G.R. Bhatt, B. Li, Y. Yu, L. Cao, J. Hone, M. Lipson, "Giant electro-refractive modulation of monolayer WS2 embedded in photonic structures" **CLEO**: Science and Innovations, STu4N. 7 (2018).
9. R. Maiti, C. Patil, R. Hemnani, M. Miscuglio, R. Amin, Z. Ma, R. Chaudhary, A. T. C. Johnson, L. Bartels, R. Agarwal, V. J. Sorger, "Loss and Coupling Tuning via Heterogeneous Integration of $MoS_2$ Layers in Silicon Photonics" *Optics Materials Express,* **9**, 2, 751-759 (2018).
10. R. A. Hemnani, C. Carfano, J. P. Tischler, M. H. Tahersima, R. Maiti, L. Bartels, R. Agarwal, V. J. Sorger, "Towards a 2D Printer: A Deterministic Cross Contamination-free Transfer Method for Atomically Layered Materials", *2D Materials*: 6, 015006 (2018).
11. R. Maiti, R. A. Hemnani, R. Amin, Z. Ma, M.Tahersima, T. A. Empante, H. Dalir, R. Agarwal, L. Bartels, V. J. Sorger, "A semi-empirical integrated microring cavity approach for 2D material optical index identification at 1.55 um" *Nanophotonics* **8(3),** 435-441 (2019)**.**
12. S. Sun, A. Badaway, V. Narayana, T. El-Ghazawi, V. J. Sorger "Photonic-Plasmonic Hybrid Interconnects: Efficient Links with Low latency, Energy and Footprint" *IEEE Photonics Journal* 7, 6 (2015).
13. S. Sun, R. Zhang, J. Peng, V. K. Narayana, H. Dalir, T. El-Ghazawi, V. J. Sorger "MODetector (MOD): A Dual-Function Transceiver for Optical Communication On-Chip" *Optics Express* 57, 18, 130-140 (2018).
14. V. K. Narayana, S. Sun, A.-H. Badawya, V. J. Sorger, T. El-Ghazawi "MorphoNoC: Exploring the Design Space of a Configurable Hybrid NoC using Nanophotonics*" Microprocessors and Microsystems* 50, 113-126. (2017).
15. J. Meng, M. Miscuglio, J. K. George, V. J. Sorger "Electronic Bottleneck Suppression in Next-generation Networks with Integrated Photonic Digital-to-analog Converters"*arXiv* preprint: 1911:02511 (2019).
16. S. Sun, V. K. Narayana, I. Sarpkaya, J. Crandall, R. A. Soref, H. Dalir, T. El-Ghazawi, V. J. Sorger "Hybrid Photonic-Plasmonic Non-blocking Broadband 5×5 Router for Optical Networks"*IEEE Photonics Journal* 10, 2 (2018).
17. J. Peng, S. Sun, V. Narayana, V.J. Sorger, T. El-Ghazawi "Integrated Nanophotonics Arithmetic for Residue Number System Arithmetic"*Optics Letters* 43, 9, 2026-2029 (2018).
18. S. Sun, M. Miscuglio, R. Zhang, Z.Ma, E. Kayraklioglu, C. Chen, J. Crandall, J. Anderson, Y. Alkabani, T. El-Ghazawi, V. J. Sorger "Analog Photonic Computing Engine as Approximate Partial Differential Equation Solver"*arXiv* preprint: 1911.00975 (2019).
19. M. Miscuglio, A. Mehrabian, Z. Hu, S.I. Azzam, J.K. George, A.V. Kildishev, M. Pelton, V.J. Sorger "All-optical Nonlinear Activation Function for Photonic Neural Networks" *Optical Material Express* 8(12), 3851-3863 (2018).
20. J. K. George, A. Mehrabian, R. Armin, J. Meng, T. Ferreira De Lima, A. N. Tait, B. Shastri, P. Prucnal, T. El-Ghazawi, V. J. Sorger "Noise and Nonlinearity of Electro-optic Activation Functions in Neuromorphic Compute Systems" *Optics Express* 27, 4 (2019).

21. A. Mehrabian, M. Miscuglio, Y. Alkabani, V. J. Sorger, T. El- Ghazawi "A Winograd-based Integrated Photonics Accelerator for Convolutional Neural Networks" ***IEEE J. of Selected Topics in Quantum Electronics*** 26(1), 1-12 (2019).
22. M. Miscuglio, G.C. Adam, D. Kuzum, V.J. Sorger "Roadmap on Material-Function Mapping for Photonic-Electronic Hybrid Neural Networks" ***APL Materials*** 7, 100903 (2019)
23. Y. Alkabani, M. Miscuglio, V.J. Sorger, T. El-Ghazawi "OE-CAM: A Hybrid Opto-Electronic Content Addressable Memory" ***arXiv*** preprint: 1912:02221 (2019).
24. R. Amin, R. Maiti, J. K. George, X. Ma, Z. Ma, H. Dalir, M. Miscuglio, V. J. Sorger "A lateral MOS-Capacitor Enabled ITO Mach- Zehnder Modulator for Beam Steering" ***Journal Lightwave Technology*** doi: 10.1109/JLT.2019.2956719 (*2019*).
25. C. Ye, S. Khan, Z.R. Li, E. Simsek, V. J. Sorger "λ-Size ITO and Graphene-based Electro-optic Modulators on SOI" ***IEEE Selected Topics in Quantum Electronics***, 4, **20 (*2014*).**
26. V. J. Sorger, R. Amin, J. B. Khurgin, Z. Ma, S. Khan "Scaling Vectors for Atto-Joule per Bit Modulators" ***Journal Optics*** 20, 014012 (*2018*).
27. K. Liu, S. Sun, A. Majumdar, V. J. Sorger, "Fundamental Scaling Laws in Nanophotonics" ***Nature Scientific Reports*** 6, 37419 (*2016*).
28. R. Amin, M. Tahersima, Z. Ma, C. Suer, K. Liu, H. Dalir, V. J. Sorger "Low-loss Tunable 1-D ITO-slot Photonic Crystal Nanobeam Cavity" ***Journal Optics*** 20, 054003 (*2018*).